\documentstyle[epsf,psfig]{mn}

\def\K{{\rm\thinspace K}}

\def\kpc{{\rm\thinspace kpc}}

\def\Mpc{{\rm\thinspace Mpc}}
\def\Msun{\hbox{$\rm\thinspace M_{\odot}$}}

\def\yr{{\rm\thinspace yr}}

\def\h50{\hbox{$\rm\thinspace h_{50}$}}
\def\h50m1{\hbox{$\rm\thinspace h_{50}^{-1}$}}

\def\Msun{\mathop{\rm M_{\odot}}\nolimits}

\def\Mpc{\mathop{\rm Mpc}\nolimits}
\def\kpc{\mathop{\rm kpc}\nolimits}

\def\K{\mathop{\rm K}\nolimits}

\def\etal{{\it et al.\thinspace}}
\def\eg{{\it e.g.\ }}

\def\P3M{\hbox{$P^{3}M$}}

\def\AP3M{\hbox{$AdP^{3}M$}}

\def\cc2{c2}
\def\cc3{c3}
\def\cc4{c4}
\def\cc{c}

\def\aap{A\&A}
\def\aapr{A\&AR}

\def\aj{AJ}

\def\apj{ApJ}
\def\apjl{ApJL}

\def\mnras{MNRAS}
\def\nat{Nat}

\def\etal{{\it et al.\thinspace}}
\def\eg{{\it e.g.\ }}

\def\spose#1{\hbox to 0pt{#1\hss}}
\def\approxlt{\mathrel{\spose{\lower 3pt\hbox{$\sim$}}
	\raise 2.0pt\hbox{$<$}}}
\def\approxgt{\mathrel{\spose{\lower 3pt\hbox{$\sim$}}
	\raise 2.0pt\hbox{$>$}}}

\def\<{\thinspace}

\def\boxit#1{\vbox{\hrule\hbox{\vrule\kern3pt\vbox{\kern3pt
          #1 \kern3pt}\kern3pt\vrule}\hrule}}

\title[The X-ray properties of galaxy clusters]{The effect of radiative cooling
on the X-ray properties of galaxy clusters.}

\author[F. R. Pearce \etal]
  {F.~R.~Pearce$^{1,}$\thanks{email:F.R.Pearce@durham.ac.uk}, 
  P.~A.~Thomas$^2$, H.~M.~P.~Couchman$^3$ \&
  A.~C.~Edge$^{1}$ \\
  $^1$Department of Physics, University of Durham, Durham, DH1 3LE, UK\\
  $^2$Astronomy Centre, CPES, University of Sussex, Falmer, Brighton,
  BN1 9QJ, UK\\
  $^3$Department of Physics and Astronomy, McMaster University,
  Hamilton, Ontario, L8S 4M1, Canada
}
\date{Accepted 1999 January 00.
      Received 1999 January 00;
      in original form 1999 January 00}

\pubyear{1999}
\begin{document}
\maketitle
\begin{abstract} 

In this paper, we investigate the effect of cooling on the X-ray
properties of galaxy clusters.  We have performed N-body,
hydrodynamical simulations both with and without the effects of
radiative cooling, but neglecting the effects of star formation and
feedback.
We show that radiative cooling produces an inflow of high-entropy gas
from the outer parts of the cluster, thus \emph{raising} the cluster
temperature and \emph{decreasing} the X-ray luminosity.  With
radiative cooling clusters are on average three to five times less
luminous in X-rays than the same cluster simulated without cooling.
However, we do not produce a large constant-density core in either the
gas or the dark matter distributions.
Our results contradict previous work in which cooling raises the X-ray
luminosity and deposits an unreasonably large amount of mass in the
central cluster galaxy.  We achieve this by selecting our numerical
resolution in such a way that a reasonable fraction of the baryonic
material cools and by decoupling the
hot and cold gas in our simulations, a first step towards modelling
multiphase gas. We emphasise that globally cooling a sensible amount
of material is vital and the presence or absence of massive central
concentrations of cold baryonic material has a dramatic effect upon
the resultant X-ray properties of the clusters.

\end{abstract}


\section{Introduction}

Clusters of galaxies are the largest virialised structures in the
Universe, evolving rapidly at recent times in many popular
cosmological models. Even at moderate
redshifts the number of large dark matter halos in a cold dark matter
Universe with a significant, positive cosmological constant is higher
than in a standard cold dark matter Universe and it is precisely
because both the number density and size of large dark matter halos
evolve at different rates in popular cosmological models that
observations of galaxy clusters provide an important discriminator
between rival cosmologies.

The advent of X-ray satellites opened up a whole new area in
observational astronomy. Hot gas, typically at temperatures of
$10^{7-8}\K$ sitting in the deep potential wells of galaxy clusters emits
radiation via thermal bremsstrahlung.  This emission is heavily biased
towards the central regions of the cluster because the flux is
weighted as the gas density squared.  
Given the gas temperature and the X-ray surface brightness, the gas
column density and a spherically symmetric gas density profile can be
estimated.  If the hot gas is assumed to reside close to hydrostatic
equilibrium within the dark matter halo the underlying dark matter
density distribution can be derived.



Since the early work of Abramopoulos \& Ku (1983) and Jones \& Forman
(1984) on the radial density profiles of large galaxy clusters, debate
has raged about the presence or absence of large, $ \sim
250\,h^{-1}\kpc$, constant density cores in the X-ray emitting
gas. Unfortunately X-ray imaging is notoriously difficult because of
the inherent large beam size.  In addition, the centre of the emission
region is not easy to determine and any centering error also acts to
smooth out any central increase when the results are azimuthally
averaged and plotted as a radial profile (Beers \& Tonry 1986).  More
recent high resolution images of several clusters have helped to
resolve this controversy. Some galaxy clusters do indeed appear to
exhibit a large, resolved core but others have a much smaller core,
close to the resolution threshold of the instrument. Both White, Jones
\& Forman (1997) (who studied 207 clusters imaged by the Einstein
observatory) and Peres \etal (1998) (who looked at ROSAT observations
of the flux-limited sample of clusters provided by Edge \etal 1990)
suggest that this dichotomy relies on the presence or absence of a
cooling flow. Clusters with a cooling flow appear to have small cores
(around $50h^{-1}\kpc$) whilst clusters without cooling flows have
much larger cores. The work of Allen (1998) on the discrepancy between
the large X-ray core radii and the small core radii deduced from
strong lensing observations (e.g. Kneib \etal 1996) also reached the
conclusion that cooling flow type clusters had small core radii in
their matter distributions.  Because the X-ray flux rises as the local
density squared, the total X-ray emission from a cluster is very
sensitive to the central density.

High resolution N-body simulations of galaxy clusters (Moore \etal
1998) produce a radial dark matter profile that has no core.  The
radial profile continues to rise until the resolution threshold is
reached, well within the required radius if the gas is to trace the
dark matter and still reproduce the X-ray observations. The production
of a central constant density dark matter core has been a long
standing problem for collisionless dark matter simulations (Pearce,
Thomas \& Couchman 1993, Navarro \& White 1994), although in previous
work the resolution threshold was still close to the observed core
sizes and so until the latest high resolution studies this was still a
tentative result.

Recently several groups have tried to reconstruct the radial
temperature profile of galaxy clusters. Markevitch \etal (1998) used
ASCA data from 30 clusters and concluded that the temperature falls
steeply at large radii.  However, Irwin, Bregman \& Evrard (1999)
analysed ROSAT PSPC images of 26 clusters and concluded that the
radial temperature profiles were generally flat out to the virial
radius.

The inclusion of a gaseous component into simulations allows the
previously assumed relationship between the gas and the dark matter to
be derived directly. The first study of this type (Evrard 1990) was
carried out without the effects of radiative cooling but reproduced
well many theoretical predictions such as a bias between the dark
matter and the baryonic material. Similar non-cooling simulations have
proved popular (Cen \& Ostriker 1994, Kang \etal 1994, Bryan \etal
1994, Navarro, Frenk \& White 1995, Bartelmann \& Steinmetz 1996, Eke,
Navarro \& Frenk 1998, Bryan \& Norman 1998) and such a simulation
formed the basis of the Santa Barbara project in which a dozen groups
simulated the formation of the same galaxy cluster (Frenk \etal 1999).

Extending this work to include a dissipative component has proved
difficult because the formation of galaxies introduces many additional
physical effects. Metzler \& Evrard (1994) and Evrard, Metzler \&
Navarro (1996) circumvented this by introducing a galactic component
by hand into a simulation that didn't follow radiative cooling of the
gas to study the effects of feedback on the cluster profile. Fully
consistent attempts to follow the radiative cooling of the hot
intra-cluster gas have only recently been achievable because of the
extra computational overhead involved. Early attempts to include the
effects of radiative cooling (Thomas \& Couchman 1992, Katz \& White
1993, Evrard, Summers \& Davis 1994, Frenk \etal 1996) either suffered
from poor resolution, focussed on the galactic population or suffered
from overmerging effects.

In a now classic paper Katz \& White (1993) examined the effect of
radiative cooling on the X-ray profile of a single galaxy cluster, a
study repeated recently by Lewis \etal (1999). Their simulated cluster
has properties that are not observed; 
a Virgo sized cluster with a super-massive central
galaxy and an enormous $400\Msun/\yr$ associated cooling flow. As they
point out, massive brightest cluster galaxies of this size are
observed within the Universe, as are massive cooling flows, however
they fail to stress their rarity, particularly in objects of similar
size to the Virgo cluster.  More recently Suginohara \& Ostriker
(1998) simulated a different cluster and also produced an object which
has properties they themselves admit are unobserved -- ``The high resolution
simulation resulted in a gas density profile steeply rising toward the
center, with consequent very high X-ray luminosity; however, these
properties are not observed''. They suggest that feedback of energy
from supernovae might account for the discrepancy.  In this paper, we
obtain quite different results because, unlike previous studies, a realistic
fraction of the baryonic material has cooled to form galaxies. Large,
bright central cluster galaxies can have a dramatic
effect on the cluster potential and consequently the X-ray properties.
A reasonable treatment of these objects is extremely important in studies
of this type. To prevent too much material cooling to form the central
object within each halo we employ a modified form
of smoothed-particle hydrodynamics --- this will be
described further in Section~2, below.

If the effects of cooling are included in the models then one might
expect the entropy of the intracluster medium to decrease.
Paradoxically, several mechanisms have been suggested that may produce
large, constant-density gas cores by \emph{raising} the entropy of the
gas at the centre of the cluster:
\begin{itemize}
\item Radiative cooling is very efficient in small dark matter halos because
the cooling time is less than the dynamical time (White \& Rees
1978). These knots of cold, dense gas can be equated with
proto-galaxies and, as they are dense and collapsing, they may be
reasonably expected to produce stars (Katz 1992). Star formation leads
to energy feedback into the interstellar medium via supernovae
explosions (Katz 1992, Mihos \& Hernquist 1994, Navarro \& White 1994,
Gerritsen \& Icke 1997). Unless the gas immediately recools, this
heating acts to increase the entropy of the surrounding material,
pushing it onto a higher adiabat and preventing it settling to the
very high densities and temperatures required for it to trace the
underlying dark matter (Wu, Fabian \& Nulsen 1998).
\item The presence of galaxies orbiting within the cluster potential
acts to stir up the gas, heating it as friction and turbulence
dissipate the galaxies' velocity, simultaneously producing velocity and
spatial bias in the galaxy distribution (Frenk \etal 1996).  This
effect is most pronounced in the centre of the cluster.
\item A third mechanism for producing a core is that radiative
cooling of the gas at the centre of each potential well acts as a
drain on the low entropy material (which cools preferentially).  If
the remaining gas cannot cool rapidly enough, a core would develop
because only high entropy material remains, an effect postulated in
the first paper to include radiative cooling (Thomas \&
Couchman 1992) and later reiterated by Waxman \& Miralda-Escude (1995)
and Bower (1997).  It is this mechanism that we investigate in this
paper.  We show that the entropy of the intracluster medium is indeed
increased, that this leads to a greatly reduced X-ray luminosity, but
that it does \emph{not} give a large, constant-density core.
\end{itemize}

The remainder of this paper is laid out as follows: in Section~2 we
present the large hydrodynamical simulations, both with and without
the effects of radiative cooling, that we have performed; in Section~3
we extract radial density and temperature profiles for the 20 largest
galaxy clusters within each simulation and contrast the profiles with
the underlying dark matter distribution; this is followed in Section~4
by a discussion of our findings.

\section{The simulations}

The simulations that we have carried out use the adaptive
particle-particle, particle-mesh (AP$^3$M) method (Couchman 1991)
coupled to the smoothed particle hydrodynamics (SPH) technique
(Gingold \& Monaghan 1977, Lucy 1977) to follow 2 million gas and 2
million dark matter particles in a box of side $100\Mpc$ (Couchman,
Thomas \& Pearce 1995, Pearce \& Couchman 1997).  

We have performed 
simulations in two types of flat cold dark matter cosmology, one
standard (SCDM) and one with a cosmological constant ($\Lambda$CDM),
with the same parameters assumed by Jenkins
\etal (1998) ($\Omega=1.0$, $\Lambda=0.0$, h=0.5, $\sigma_8=0.6$ for
the former and $\Omega=0.3$, $\Lambda=0.7$, h=0.7, $\sigma_8=0.9$ for
the latter).  The baryon fraction was set from Big Bang
nucleosynthesis constraints, $\Omega_bh^2=0.015$ (Copi, Schramm \&
Turner 1995) and we have assumed an unevolving gas metallicity of 0.3
times the solar value.  These parameters produce a gas mass per
particle of $2\times10^9\Msun$ in each case and are summarised in
Table~1.  The dark matter mass is only slightly lower than that given
by Steinmetz \& White (1997---their Equation~9, but note that there is
a typographical error) at which artificial 2-body heating balances
radiative cooling.  Thus we expect there to be some numerical heating
in our simulations.  However, as we have chosen to neglect real heat
sources, such as supernovae in galaxies, this is of little importance
and does not affect our conclusions which concern the differences
between runs with and without radiative cooling.

Since we smooth over 32 SPH particles, the smallest gaseous object that
can be effectively resolved has a mass of $6.4\times10^{10}\Msun$.  We
employ a comoving $\beta$-spline gravitational softening equivalent to
a Plummer softening of $10h^{-1}\kpc$ for redshifts $0<z<1.5$ (2.5 for
$\Lambda$CDM)---at earlier times the softening has a fixed physical size,
with the minimum SPH resolution set to match this.  We note that a
spatial resolution of $10h^{-1}\kpc$ is over twice the typical scale
length of elliptical galaxies and this may lead to enhanced tidal
disruption, drag and merging within the largest clusters of objects.
However, the force softening cannot be reduced further without introducing
2-body effects. A smaller softening would also lead to a further
increase in the number of timesteps required; we already require
around 10000 for each cooling run.  

In addition to these two simulations which both included the effects
of radiative cooling, we repeated the $\Lambda$CDM model without
cooling. This simulation has input parameters close to those used by
Eke, Navarro \& Frenk (1998) who used the same cosmology and the
resultant clusters look very similar and exhibit similar X-ray
properties (see Fig.~\ref{fig.lxtall} for a comparison).

\begin{table}
\begin{tabular}{lcc}
Cosmology	 	& $\Lambda$CDM 	& SCDM \\
$\Omega$		& 0.3		& 1.0 \\
$\Lambda$		& 0.7		& 0.0 \\
$\Omega_b$		& 0.03		& 0.06 \\
$\sigma_8$		& 0.9		& 0.6 \\
$h$			& 0.7		& 0.5 \\
boxsize ($h^{-1}\Mpc$)	& 70		& 50 \\
$M_{dm}$ ($h^{-1}\Msun$)	&$1.4\times10^{10}$& $1.6\times10^{10}$ \\
$M_{gas}$ ($h^{-1}\Msun$)	& $1.4\times10^9$	& $1.0\times10^9$ \\
soft ($h^{-1}\kpc$)		& 10.		& 10. \\
$Z_{met}$ (solar)	& 0.3 		& 0.3 \\
\end{tabular}
\caption{The main parameters for each cosmology. The parameters for
the cooling and non-cooling runs were identical.}
\end{table}

The properties of the galaxies in the two simulations with radiative
cooling have been described in Pearce \etal (1999, 2000).  They clearly
demonstrate an acceptable match to both the spatial and
luminosity distribution of observed galaxies.  This was achieved by
employing three numerical approximations: a mass-resolution
of $6.4\times10^{10}\Msun$ below which objects cannot cool
efficiently, a length-resolution (ie softening) of $10h^{-1}\kpc$, and
decoupling of the hot, halo gas from the cold galactic gas.
Improving the mass and/or length resolution would increase the
fraction of cold gas in our simulations, producing galaxies that were
too luminous to match the observations.  This would then necessitate
the introduction of feedback mechanisms that would over-complicate our
model.

Decoupling the hot halo gas is a new innovation that we feel vastly
improves the ability of SPH to model fluids in which there are large
density contrasts.  Without it, hot gas particles have their density
overestimated in the vicinity of cold gas and too much material cools
to form the central galaxy.  Normal cooling of the intracluster medium
at temperatures above $10^5$K is still handled correctly and
galaxy-galaxy mergers and viscous drag on the galaxies as they orbit
within the halo are retained. As Pearce \etal (1999) show, such a
procedure produces a set of galaxies that fit the local K-band number
counts of Gardner \etal (1997). The brightest cluster galaxies
contained within the largest halos are not excessively luminous for a
volume of this size, unlike those found in previous work (Katz \&
White 1993, Lewis \etal 1999).
The fraction of the baryonic material that cools into
galaxies within the virial radius of the large halos in our simulation
is listed in Tables 3 \& 4 and is typically around 20 percent. 
This is much less than the unphysically high value of 40
percent reported by Katz \& White (1993).  Decoupling of the hot phase
produces a galactic population and cold gas fraction that is well
matched to the observations.

\section{Results}

\subsection{Extracting objects}

For the purposes of this paper we are interested in only the largest
objects within each simulation, as only these contain sufficient
mass to produce the deep potential wells required to retain hot, X-ray
emitting gas. We centre each cluster on 
the peak of the hot gas density, a position
that coincides with the centre of the X-ray emission. 
This prevents the introduction of an
artificial constant-density core which may arise with any other
choice of centre --- for clusters with significant
substructure the centre-of-mass can lie a long way from the
centre of the X-ray emission.

The virial radius for each of our clusters was defined as the
spherical region, surrounding each cluster centre, that enclosed an
overdensity of 178 for SCDM and 324 for $\Lambda$CDM (Eke, Cole \& Frenk 1996).
Each catalogue was then cleaned by ordering it in size and deleting
the smaller of any overlapping clusters.  The 20 most massive clusters
in each of the catalogues were then used for the work presented here.
The properties of the clusters are presented in Tables~2--4.  The 16
largest clusters found in the non-cooling simulation are found in the
list of the 20 largest clusters in the other two models.  The index of
the matching cluster from the non-cooling run is given in Tables~3 \&
4.

Each of the extracted clusters was checked for substructure by
comparing the centre of the X-ray emission to the median position of
the particles within the virial radius, a statistic that has been
shown to be a useful indicator of the presence of substructure by
Thomas \etal (1998).  The results of this test are shown
in Tables~2--4.   All the clusters with an offset of more than 7
percent of the virial radius (6 in each case) were noted and are shown
as dotted lines on Figures~\ref{fig.dmprof}, \ref{fig.entprof},
\ref{fig.denprof}, \ref{fig.tempprof} \& \ref{fig.lxprof} and as open
symbols on Figure~\ref{fig.lxtall}.  

\begin{table*}
\begin{tabular}{lcccccccc}
Run & ${\rm M_v}$ & Virial rad & substruct & $N_{dm}$ & $N_{hotgas}$ & $T_X$ &
$L_X$ \\
$\Lambda$nc1  & 8.54& 1.97& 0.023& 63175& 54842 & 6.31&  49.3\\
$\Lambda$nc2  & 3.07& 1.40& 0.086& 22643& 18175 & 2.61&  3.43\\
$\Lambda$nc3  & 2.54& 1.31& 0.071& 18770& 16094 & 2.04&  3.39\\
$\Lambda$nc4  & 1.82& 1.17& 0.097& 13431& 11218 & 1.55&  2.32\\
$\Lambda$nc5  & 1.53& 1.10& 0.073& 11256&  9642 & 1.92&  4.02\\
$\Lambda$nc6  & 1.53& 1.10& 0.033& 11235&  9072 & 1.71&  2.64\\
$\Lambda$nc7  & 1.43& 1.08& 0.013& 10552&  9262 & 2.12&  4.26\\
$\Lambda$nc8  & 1.24& 1.03& 0.018&  9138&  7780 & 1.89&  2.96\\
$\Lambda$nc9  & 1.19& 1.01& 0.083&  8766&  7201 & 1.44&  0.69\\
$\Lambda$nc10 & 1.10& 0.99& 0.018&  8091&  6414 & 1.49&  1.63\\
$\Lambda$nc11 & 1.09& 0.99& 0.032&  8054&  7204 & 1.84&  2.44\\
$\Lambda$nc12 & 1.03& 0.96& 0.020&  7573&  6250 & 1.59&  3.09\\
$\Lambda$nc13 & 0.99& 0.96& 0.008&  7351&  6187 & 1.97&  3.33\\
$\Lambda$nc14 & 0.98& 0.95& 0.006&  7227&  5956 & 1.65&  1.43\\
$\Lambda$nc15 & 0.92& 0.93& 0.012&  6829&  5542 & 1.43&  0.98\\
$\Lambda$nc16 & 0.83& 0.90& 0.008&  6169&  4958 & 1.47&  1.34\\
$\Lambda$nc17 & 0.83& 0.90& 0.241&  6130&  4740 & 0.91&  0.47\\
$\Lambda$nc18 & 0.83& 0.90& 0.035&  6082&  5038 & 1.36&  1.14\\
$\Lambda$nc19 & 0.78& 0.88& 0.030&  5804&  5321 & 1.41&  1.01\\
$\Lambda$nc20 & 0.76& 0.87& 0.027&  5636&  4456 & 1.27&  1.37\\
					
\end{tabular}
\caption{Data for the 20 largest clusters (by mass) in the 
$\Lambda$CDM run without cooling. Quoted are the cluster number,
the virial mass in units
of $10^{14}h^{-1}\Msun$, the virial radius in $h^{-1}\Mpc$, the value of the
substructure parameter defined in the text (a larger number implies
more substructure), the number of
dark matter particles within the virial radius, the number of gas particles
within this radius,
the emission weighted mean temperature (in keV/$k$) and the bolometric
X-ray luminosity in units of $10^{43}$erg/s/$h^2$}.
\end{table*}

\begin{table*}
\begin{tabular}{lcccccccccccc}
Cluster & Match & ${\rm M_v}$ & Virial rad & substruct &$N_{dm}$ & $N_{hotgas}$
& $N_{coldgas}$ & $N_{gal}$ & cold fraction & $T_X$ &
$L_X$ \\
$\Lambda$1     &  1& 8.75& 1.98& 0.025  & 64516 & 46928& 7528 & 2894& 0.14 & 4.64  &14.2 \\
$\Lambda$2     &  2& 2.79& 1.35& 0.103	& 20612 & 13485& 3939 & 1578& 0.23 & 3.41  &1.10 \\
$\Lambda$3     &  3& 2.58& 1.31& 0.044	& 19070 & 12035& 4109 & 1319& 0.25 & 2.58  &1.04 \\
$\Lambda$4     &  4& 1.60& 1.12& 0.182	& 11855 &  7409& 2467 & 1082& 0.25 & 2.22  &0.61 \\
$\Lambda$5     &  7& 1.50& 1.10& 0.014	& 11118 &  7340& 2262 & 1253& 0.24 & 0.87  &2.75 \\
$\Lambda$6     &  5& 1.48& 1.09& 0.008	& 10936 &  7369& 2148 & 1338& 0.23 & 2.09  &1.30 \\
$\Lambda$7     &  6& 1.29& 1.04& 0.009	&  9526 &  6649& 1613 & 1241& 0.20 & 2.50  &1.28 \\
$\Lambda$8     & --& 1.22& 1.02& 0.020	&  8978 &  4901& 2500 & 1074& 0.34 & 1.59  &0.21 \\
$\Lambda$9     &  8& 1.21& 1.02& 0.049	&  8916 &  5252& 2137 & 1182& 0.29 & 2.57  &0.29 \\
$\Lambda$10    & 10& 1.14& 1.00& 0.015	&  8397 &  5345& 1764 &  773& 0.25 & 1.97  &0.35 \\
$\Lambda$11    &  9& 1.11& 0.99& 0.038	&  8213 &  4573& 1770 & 1172& 0.28 & 1.91  &0.23 \\
$\Lambda$12    & 13& 0.98& 0.95& 0.004	&  7231 &  5097& 1232 & 1031& 0.19 & 2.18  &0.82 \\
$\Lambda$13    & 12& 0.98& 0.95& 0.002	&  7213 &  4630& 1327 &  979& 0.22 & 1.94  &0.40 \\
$\Lambda$14    & 11& 0.96& 0.94& 0.050	&  7074 &  4154& 1413 & 1041& 0.25 & 1.73  &0.23 \\
$\Lambda$15    & 14& 0.92& 0.93& 0.189	&  6836 &  4359& 1602 &  648& 0.27 & 1.74  &0.32 \\
$\Lambda$16    & 16& 0.92& 0.93& 0.037	&  6766 &  4156& 1533 &  807& 0.27 & 1.64  &0.27 \\
$\Lambda$17    & 18& 0.87& 0.91& 0.073	&  6386 &  3713& 1704 &  523& 0.31 & 1.03  &0.22 \\
$\Lambda$18    & --& 0.82& 0.89& 0.192	&  6022 &  3346& 1707 &  348& 0.34 & 1.34  &0.21 \\
$\Lambda$19    & 15& 0.76& 0.87& 0.043	&  5578 &  3155& 1577 &  828& 0.33 & 1.38  &0.30 \\
$\Lambda$20    & --& 0.75& 0.87& 0.179	&  5539 &  3264&  883 &  332& 0.21 & 2.00  &0.18 \\
\end{tabular}
\caption{Data for the 20 largest clusters (by mass) in the 
$\Lambda$CDM run which includes cooling. Quoted are the cluster number, the
number of the matching cluster in the  $\Lambda$CDM run without
cooling (if one exists), the virial mass in units
of $10^{14}h^{-1}\Msun$, the virial radius in $h^{-1}\Mpc$, the value of the
substructure parameter defined in the text, the number of
dark matter particles within the virial radius, the number of gas particles
above and below $12000\K$ within this radius, the number of cold
particles within $100h^{-1}\kpc$ of the centre, the fraction of the
baryonic material within the virial radius that has been able to cool,
the emission weighted mean temperature (in keV/$k$) and the bolometric
X-ray luminosity in units of $10^{43}$ergs/s/$h^2$.}
\end{table*}

\begin{table*}
\begin{tabular}{lcccccccccccc}
Cluster & Match & ${\rm M_v}$ & Virial rad & substruct & $N_{dm}$ & $N_{hotgas}$
& $N_{coldgas}$ & $N_{gal}$ & cold fraction & $T_X$ &
$L_X$ \\
S1     &  1& 9.65&1.67& 0.069	& 58750 & 50740& 3490& 1286& 0.06  &7.48 & 6.55\\
S2     &  2& 4.15&1.26& 0.144 	& 25181 & 18881& 2494&  805& 0.12  &4.47 & 1.09\\
S3     &  3& 2.83&1.11& 0.234 	& 17171 & 13385& 1942&  522& 0.13  &2.91 & 0.62\\
S4     &  4& 2.05&0.99& 0.071 	& 12452 &  9458& 1304&  475& 0.12  &2.27 & 0.43\\
S5     &  9& 1.89&0.97& 0.082 	& 11467 &  8775& 1408&  461& 0.14  &2.42 & 0.36\\
S6     &  5& 1.85&0.96& 0.038 	& 11221 &  8489& 1201&  631& 0.12  &2.10 & 1.07\\
S7     &  6& 1.59&0.91& 0.014 	&  9653 &  6782& 1373&  480& 0.17  &2.38 & 0.19\\
S8     &  8& 1.49&0.89& 0.020 	&  9020 &  6696&  915&  683& 0.12  &1.74 & 0.43\\
S9     &  7& 1.39&0.87& 0.015 	&  8423 &  6979&  774&  666& 0.10  &2.91 & 0.63\\
S10    & --& 1.37&0.87& 0.025 	&  8319 &  5978&  966&  596& 0.14  &2.89 & 0.24\\
S11    & 11& 1.34&0.86& 0.025 	&  8128 &  5896& 1134&  600& 0.16  &2.50 & 0.21\\
S12    & 14& 1.23&0.84& 0.117 	&  7476 &  5737&  894&  384& 0.13  &2.29 & 0.31\\
S13    & 13& 1.21&0.83& 0.025 	&  7355 &  5965&  643&  466& 0.10  &3.18 & 0.31\\
S14    & 10& 1.19&0.83& 0.007 	&  7244 &  5464&  729&  442& 0.12  &2.53 & 0.21\\
S15    & 12& 1.02&0.79& 0.052 	&  6204 &  4533&  812&  244& 0.15  &1.55 & 0.14\\
S16    & --& 1.01&0.78& 0.030 	&  6116 &  4364&  862&  371& 0.16  &1.61 & 0.10\\
S17    & --& 0.97&0.77& 0.035 	&  5874 &  4423&  671&  454& 0.13  &2.26 & 0.12\\
S18    & 20& 0.93&0.76& 0.097 	&  5656 &  3878&  888&  230& 0.19  &2.68 & 0.16\\
S19    & 15& 0.86&0.74& 0.042 	&  5196 &  3639&  775&  250& 0.18  &1.66 & 0.10\\
S20    & 17& 0.86&0.74& 0.008 	&  5186 &  4184&  445&  370& 0.10  &2.16 & 0.15\\
\end{tabular}
\caption{Data for the 20 largest clusters (by mass) in the 
SCDM run which includes cooling. Quoted are the cluster number, the
number of the matching cluster in the  $\Lambda$CDM run without
cooling (if one exists), 
the virial mass in units
of $10^{14}h^{-1}\Msun$, the virial radius in $h^{-1}\Mpc$, the value of the
substructure parameter defined in the text, the number of
dark matter particles within the virial radius, the number of gas particles
above and below $12000\K$ within this radius, the number of cold
particles within $100h^{-1}\kpc$ of the centre, the fraction of the
baryonic material within the virial radius that has been able to cool,
the emission weighted mean temperature (in keV/$k$) and the bolometric
X-ray luminosity in units of $10^{43}$ergs/s/$h^2$.}
\end{table*}

\subsection{Dark matter density profiles}

\begin{figure*}
 \centering
\psfig{file=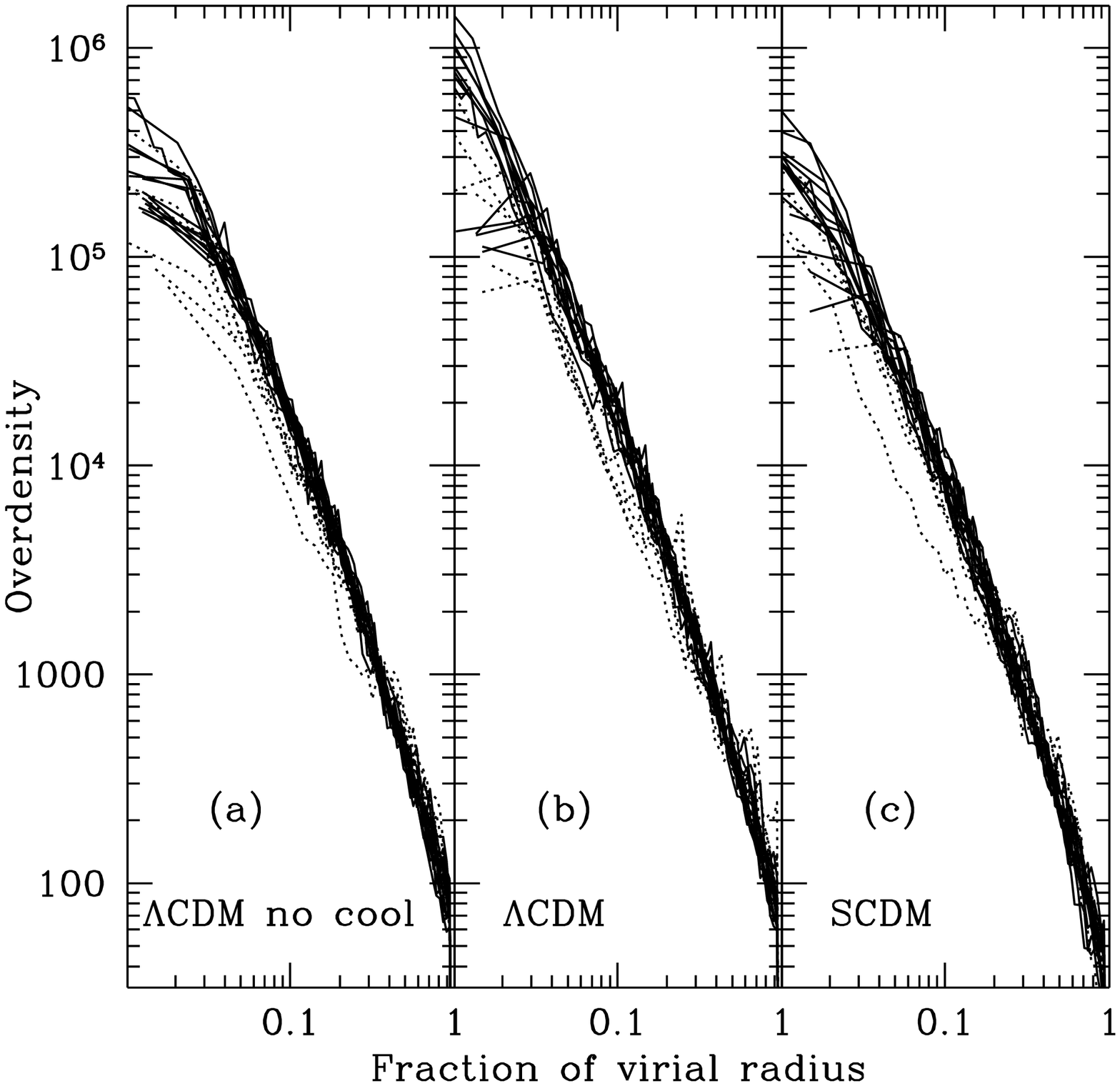,height=15cm}
\caption{The radial dark matter density profiles of the 20 largest, 
distinct dark matter halos extracted from each of the three
simulations. Plotted is the mean density within successive spherical
shells.  Each profile has been scaled to the virial radius and
overdensity relative to the cosmic mean for the dark matter.  Those
profiles marked as dotted lines are for clusters that contain
significant substructure (see text). All the profiles are started from
the point where 32 particles are enclosed.} 
\label{fig.dmprof}
\end{figure*}

The radial dark matter density profiles for the 20 largest objects in
each cosmology are shown in Figure~\ref{fig.dmprof}.  Shown is the
mean density within spherical shells, the innermost shell plotted in
the Figure containing at least 64 particles.  The dark matter profiles
of those clusters without significant substructure are similar within
each cosmology.
In the non-cooling run, the innermost bin of each density profile shows
a flattening.  This is a resolution effect: the radial extent of the
bin has to be very large in order to accommodate 64 particles.


\begin{figure*}
 \centering
\psfig{file=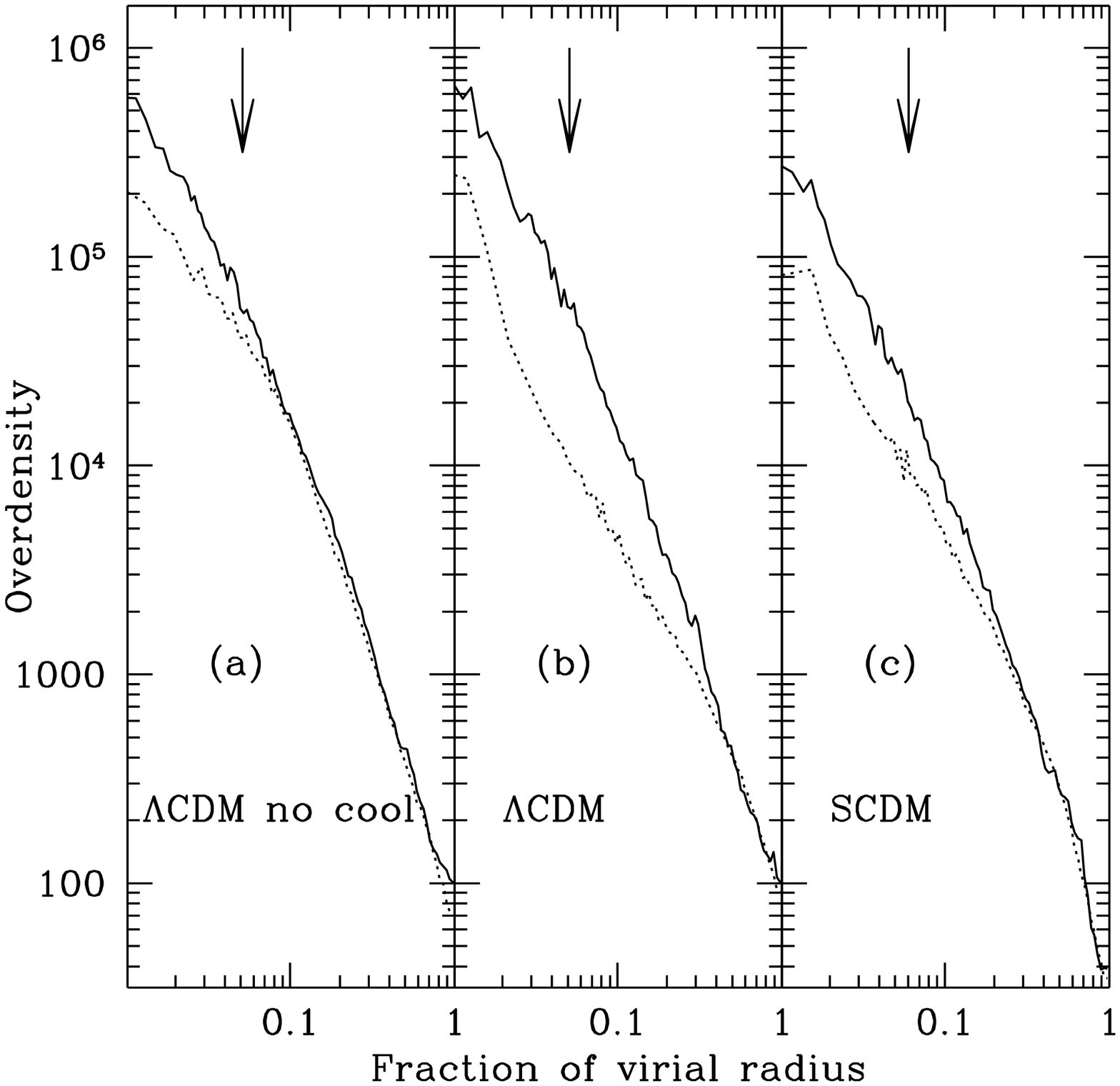,height=15cm}
\caption{The radial dark matter (solid line) and gas (dotted line) 
density profile within the virial radius of the largest 
dark matter halo extracted from each simulation. Each curve is scaled
to the virial radius of the halo and overdensity of the appropriate
phase relative to the mean cosmic density of that phase.
In all three runs there is less gas within the virial radius than dark
matter, relative to the cosmic mean of each species.
No constant density core is found in the cooling runs.
The arrow indicates a scale of 100$h^{-1}\kpc$.
}
\label{fig.nfw}
\end{figure*}

Both gas and dark matter density profiles for the largest object in
each of the runs are shown in Figure~\ref{fig.nfw}. The dark matter
profile for the $\Lambda$CDM run without cooling is reasonably well
fit by an NFW profile (Navarro, Frenk \& White 1997).  With cooling
the dark matter density profile of the clusters is not well fit by the
NFW formula, because the asymptotic slope in the central regions is
steeper than -1.
This is because, once cooling has been implemented, the large
galaxy that forms at the centre of each cluster acts to draw in more
dark matter and steepen the profile significantly in the inner
regions.
A similar effect is seen for most of the other clusters, although
in the cooling run
several show a drop in density in the innermost bin.
This indicates that the peak X-ray emissivity sometimes comes
from a galaxy that is not located at the centre of the cluster.

\subsection{Gas entropy profiles}

\begin{figure*}
 \centering
\psfig{file=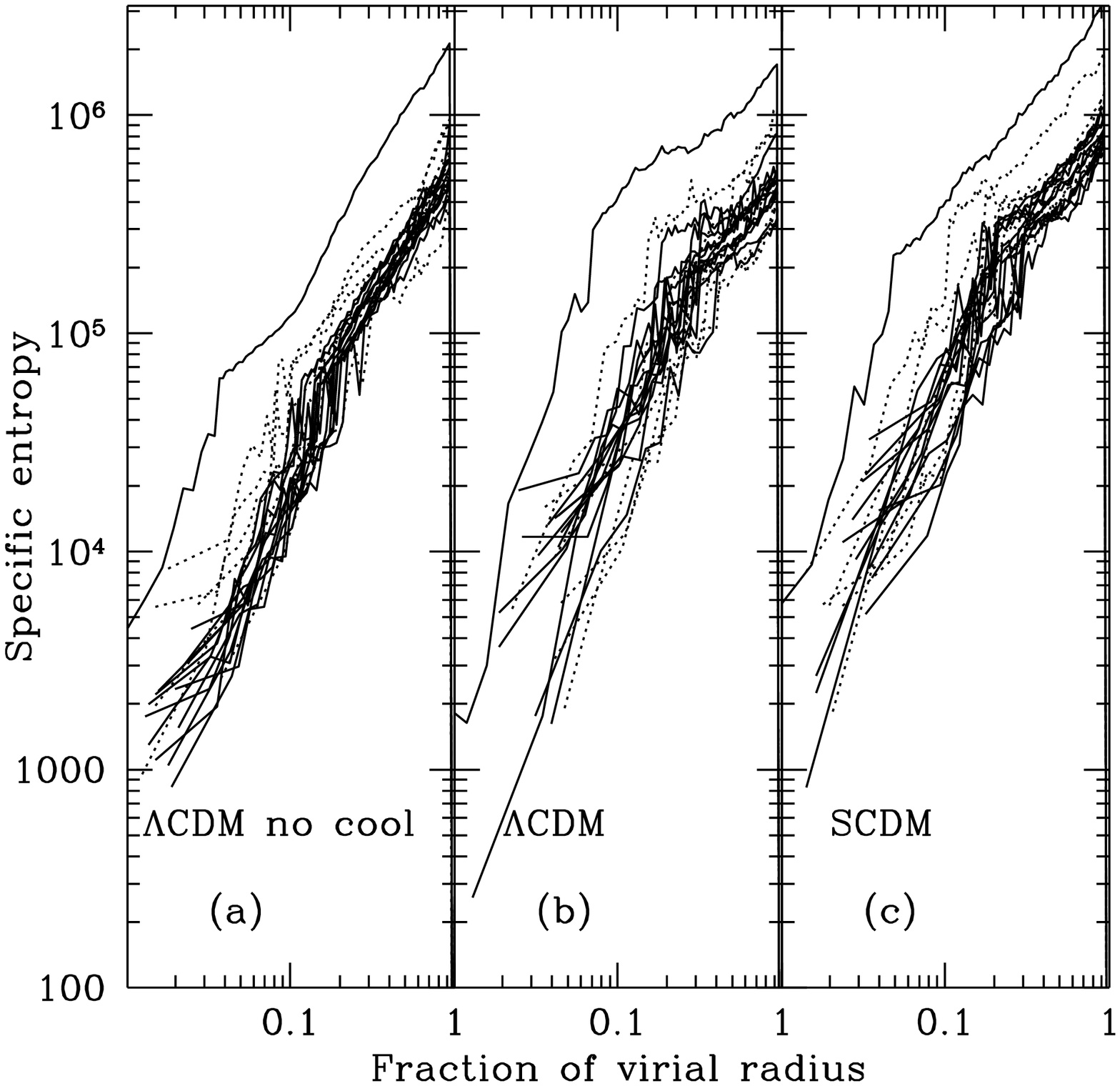,height=15cm}
\caption{The radial specific entropy profiles of the 20 largest, 
distinct halos extracted from each of the three simulations. Plotted
is the mean value of $T/n^{2/3}$ within successive spherical shells.
Each profile has been scaled to the virial radius.  Those profiles
marked as dotted lines are for clusters that contain significant
substructure (see text).}
\label{fig.entprof}
\end{figure*}

Gas entropy profiles (and also density and temperature profiles,
below) were obtained using only those particles with a temperature
exceeding $12\,000\K$.  Typical gas temperatures exceed $10^7\K$ for
halos in this mass range and we wish to exclude cold gas which lies
within galaxies or recently tidally disrupted objects.  If the cold
material were included, there would be a large density spike at the
centre of each of the clusters (which all have a central galaxy).
This object does not contribute to the X-ray emission because the
material it contains is very cold compared to the surrounding hot halo
(although the increased depth of the local potential can help to
confine dense, hot gas which can affect the bolometric X-ray
emission). The specific entropy profile is shown in Figure~\ref{fig.entprof}.
We plot the quantity $(T/\K)/(\rho/\bar\rho)^{2/3}$, where $T$ is the
temperature and $\rho$ the density, measured in units of the mean gas density,
$\bar\rho$.

Let us contrast the results for the $\Lambda$CDM
runs with and without cooling.  Firstly, note that the entropy at the
virial radii is very similar in each case---this is because cooling
has had little effect at these large radii.  Between the virial radius
and about 0.2 times the virial radius (less for the largest cluster),
the entropy profiles for the cooling run are shallower than in the
non-cooling run.  This confirms the hypothesis of Thomas \& Couchman
(1992) that cooling is able to \emph{raise} the entropy of the
intracluster medium by dragging in high-entropy material from the
outer regions of the cluster.

Within about 0.2 virial radii, the entropy profiles again steepen---it
is within this ``cooling radius'' that the cooling time is short
enough to allow significant cooling of the gas within the lifetime of
the cluster.  By the time we get to the innermost bins in the Figure,
there seems to be a spread in the entropy of the clusters in the
cooling run, with some having higher entropy and some lower entropy
than the corresponding clusters in the non-cooling run.

The SCDM run with cooling exhibits similar entropy profiles to the
$\Lambda$CDM run with cooling.

\subsection{Gas density profiles}

\begin{figure*}
\centering
\psfig{file=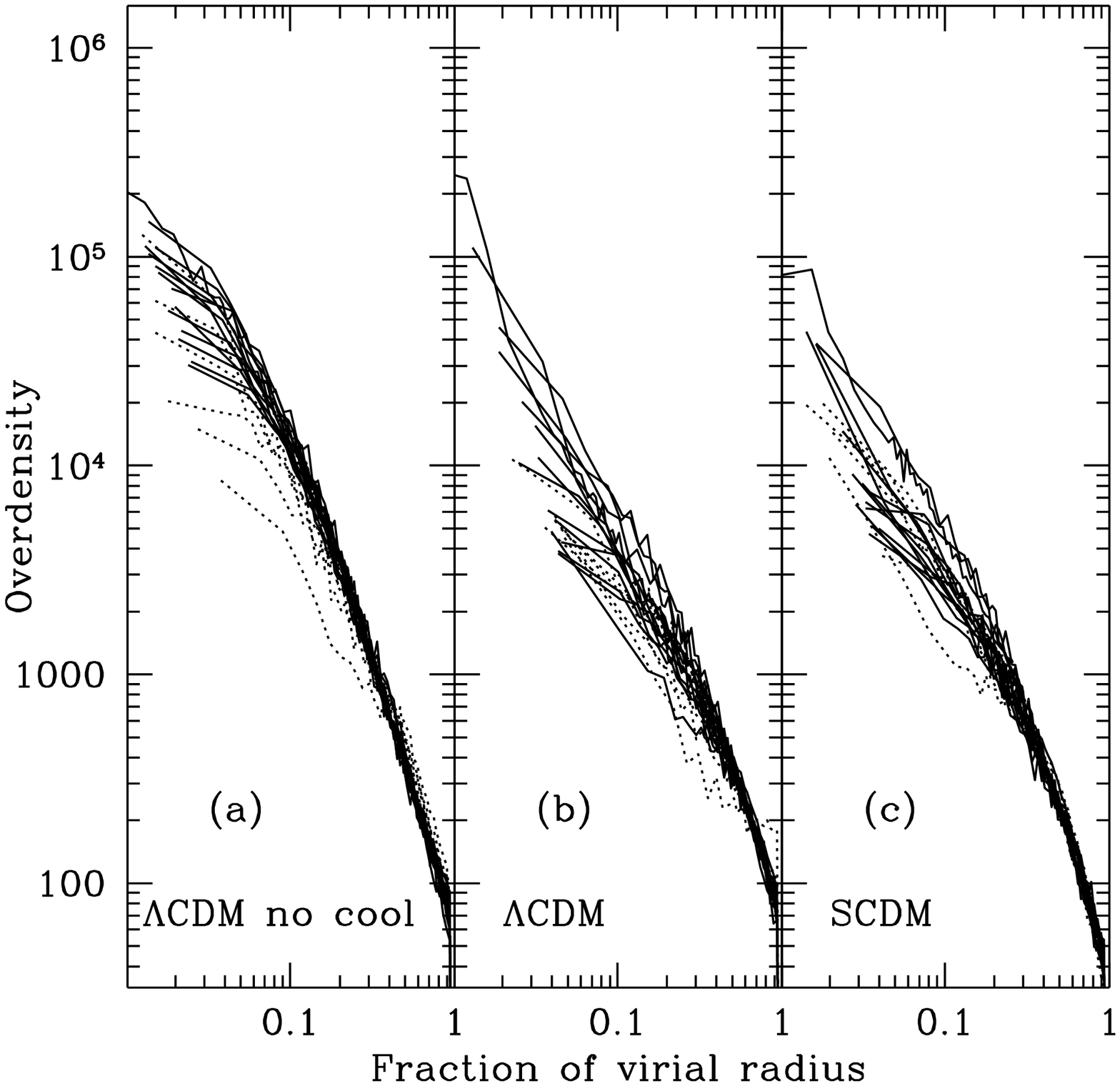,height=15cm}
\caption{The radial gas density profile of the 20 largest halos found
within each simulation.  Those halos containing significant
substructure are shown as dotted lines.  Plotted is the mean gas
density within successive spherical shells.  All the halos are scaled
to the virial radius and gas overdensity relative to the baryonic
cosmic mean.}
\label{fig.denprof}
\end{figure*}

The radial gas density profiles are displayed in
Figure~\ref{fig.denprof}.  Without cooling these profiles look very
similar to those obtained by Eke \etal (1998). Within about 0.1 virial
radii, the gas profiles are significantly shallower than the
corresponding dark matter profiles of Figure~\ref{fig.dmprof}.  This
is indicative of the fact that the hot gas has a higher specific energy
than the dark matter and, as previous authors have found 
(\eg Navarro \etal 1995, Eke \etal 1998), there is more dark matter
than gas (relative to the cosmic mean) within the
virial radius.
There is a general tendency for all the profiles to flatten
considerably in the innermost bin.  Once again, this is due to
inadequate resolution and we do not attach any significance to it.

The effect of cooling is to \emph{lower} the gas density at all radii
within the virial radius.  The suppression is greatest, a factor of
three, at about 0.1 times the virial radius, roughly corresponding to
the kink in the entropy profiles seen in
Figure~\ref{fig.entprof}.  Although the density gradients are
shallower, they do not roll over into constant-density inner core
regions.  In fact, for the larger clusters, the density continues to
rise further into the centre of the cluster than before, so that the
central density is close to that in the non-cooling case.

The profile of the largest object in both gas and dark matter for each
of the runs is shown in Figure~\ref{fig.nfw}.  The arrow indicates a
radius of 100\,$h^{-1}$kpc.  Without cooling, the gas density is
shallower than that of the dark matter within 0.1 times the virial
radius, but this inner, resolved slope of the density profile is still
$\rho \propto r^{-1}$ with no sign of a constant-density core.  As the
temperature is approximately constant within this region (see
Figure~\ref{fig.tempprof}), the X-ray luminosity is convergent and
dominated by emission from around 200\,$h^{-1}$kpc (0.1 times the
virial radius).

With cooling, the largest cluster exhibits a central density spike due
to the presence of a massive central galaxy.  This hot gas has a very
steep radial density profile, $\rho \propto r^{-3}$, and would be
classified observationally as a cooling flow of $60 h^{-2}\Msun$/yr
onto the central cluster galaxy.  Between radii of about
40\,$h^{-1}$kpc and 1\,$h^{-1}$Mpc, the density profile is a power
law, $\rho \propto r^{-1.4}$, steepening at larger radii.  Thus the
X-ray luminosity (excluding the cooling flow) comes from a much more
extended region than in the non-cooling case.

In conclusion, the gas density has been reduced by the influx of
high-entropy material, as expected.  However, this has not given rise
to constant-density inner cores.  In fact, if anything, the density
profiles now continue as a power-law closer into the centre of the
clusters.

\subsection{Radial temperature profiles}

\begin{figure*}
 \centering
\psfig{file=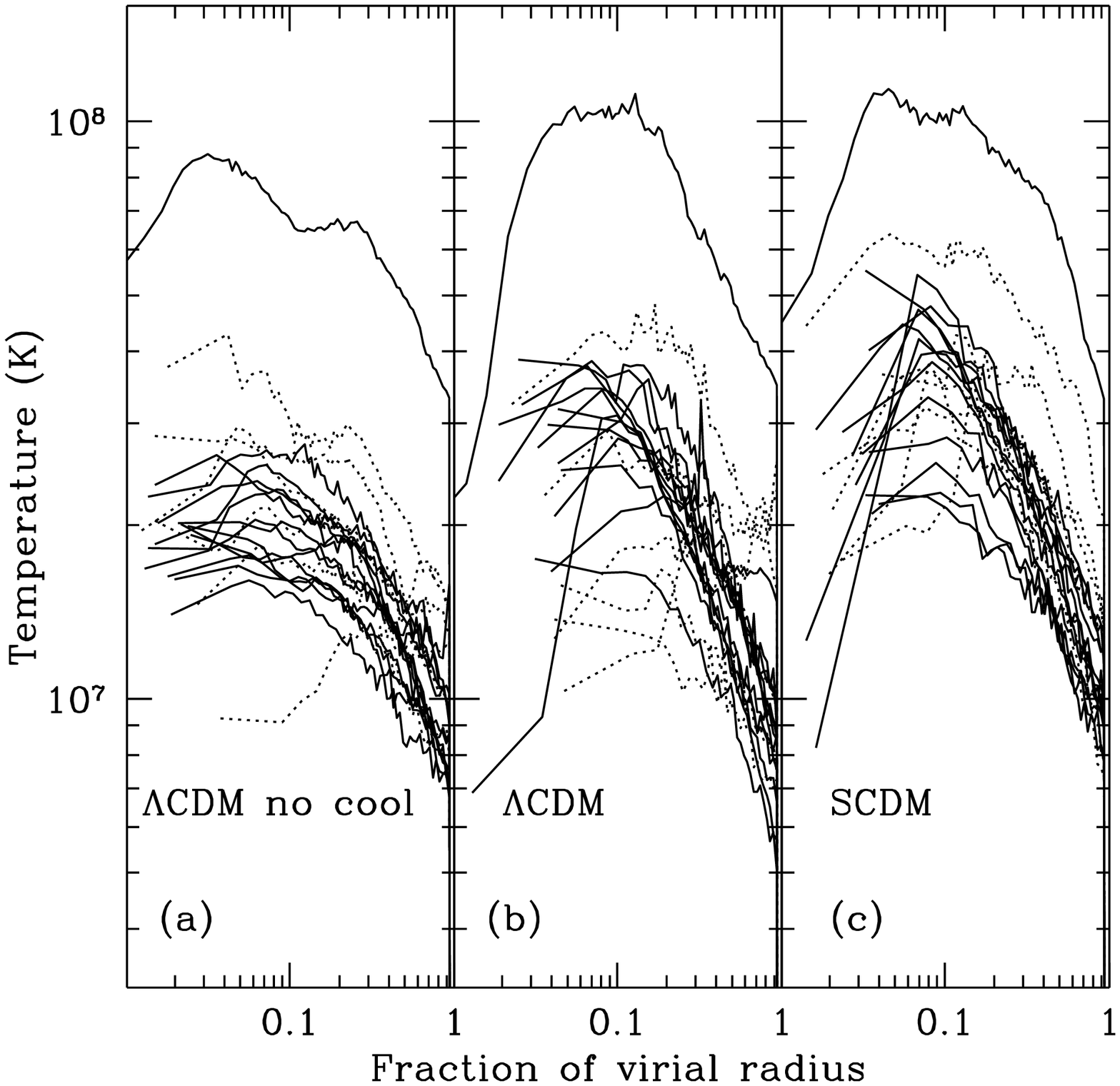,height=15cm}
\caption{The radial gas temperature profile found within each of the 20
halos extracted from each cosmology. Clusters with significant
substructure are marked as dotted lines.
All the halos are scaled to the virial radius and
temperatures (calculated as the mass-weighted 
mean temperature of the gas within successive spherical shells)
are in Kelvin. With cooling the cluster temperature reaches
a slightly higher maximum at a larger fraction of the virial radius 
whereas the temperature at the virial radius is very similar.
}
\label{fig.tempprof}
\end{figure*}

Radial temperature profiles are shown in Figure~\ref{fig.tempprof}.
The temperature profiles for the relaxed clusters from the $\Lambda$CDM
run without cooling are typical of those found in previous
work (see for example Eke \etal 1998 and references therein).  They
rise inwards from the virial radius by about a factor of two, peaking
at about 0.1 times the virial radius and then declining again, very
gradually, in the cluster centre.

Cooling makes little difference to the temperature profiles, except
that corresponding clusters in the $\Lambda$CDM runs reach a
\emph{higher} peak temperature when cooling is implemented, due to the
inflow of higher entropy gas.  The temperatures are very similar at
the virial radius, but are about 1.5 times higher at their peak than
before.  Two clusters show a precipitous decline in temperature in the
cluster centre, one of these being the largest cluster---this is
evidence for a cooling flow.

The SCDM results are very similar to those for $\Lambda$CDM.

\subsection{X-ray luminosity profiles}

\begin{figure*}
 \centering
\psfig{file=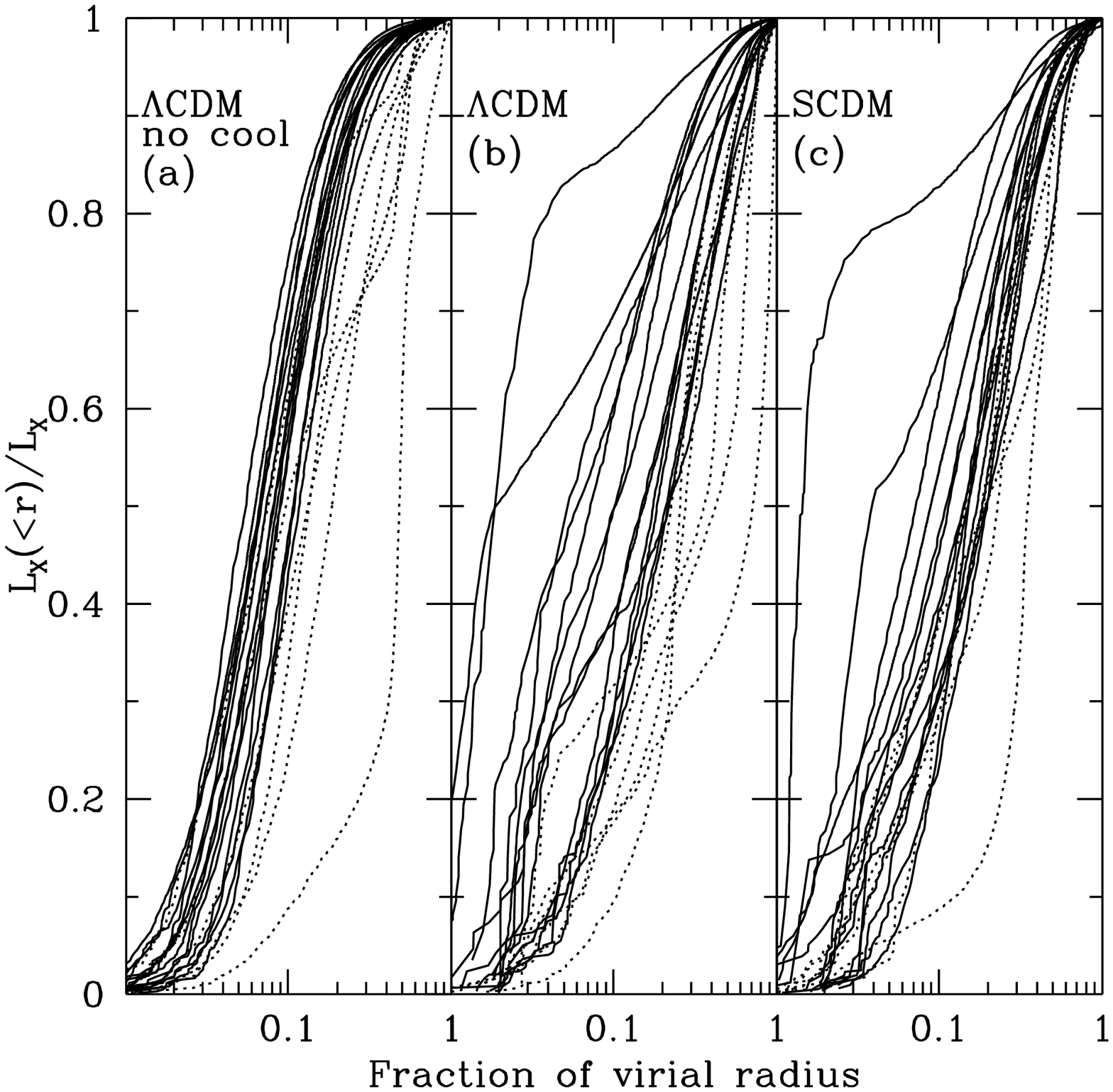,height=15cm}
\caption{The fraction of the total 
bolometric luminosity that is emitted from within the specified radius
for each of the clusters (calculated from equation~1). 
Clusters with significant substructure are shown as dotted lines.
}
\label{fig.lxprof}
\end{figure*}

We follow Navarro \etal (1995) in using the following estimator
for the bolometric X-ray luminosity of a cluster,
\begin{equation}
L_X = 4 \times 10^{32}
\sum \left(\rho_i\over\bar\rho\right)\left(T_i\over\K\right)^{1\over2}
\,{\rm erg\,s}^{-1}
\end{equation}
where the density is in units of overdensity (relative to the mean gas
density in the box, 2.86$\times10^{-31}$g/cm$^3$) and the sum extends
over all the gas particles with temperatures above $12000\K$. 

The total X-ray luminosity within the virial radius of each of the
clusters is listed in Tables~2--4.  For the simulation with cooling
the clusters are several times less luminous than those from the
corresponding non-cooling run. This contradicts the previous results of
Katz \& White (1993), Suginohara \& Ostriker (1998) and Lewis \etal
(1999) who all found the X-ray luminosity increased if cooling was
turned on.   The reason for the discrepancy is, once again, the fact
that we have decoupled the hot and cold gas, thus greatly suppressing
the cooling of the inflowing, high-entropy gas in our simulations
compared to previous ones. This causes a large reduction in the mass
of the brightest cluster galaxy compared to those produced by previous
work. Our galaxies have reasonable luminosities, mass-to-light ratios
and number counts for a volume of this size.

Note that estimates of X-ray luminosity from the non-cooling run are
not really meaningful.  A radiation rate of this magnitude can only be
sustained for a short time before depleting the intracluster medium of
gas. The cooling runs produce a more physically
self-consistent X-ray luminosity because the radiative effects are
taken into account. The X-ray
luminosity within the cooling radius is approximately equal to the
enthalpy of the gas divided by the age of the cluster.

We plot these bolometric luminosities as a function of radius for each
of our clusters in Figure~\ref{fig.lxprof}.  Clearly the relative
contribution to the total X-ray emission from different radii is very
different for the cooling and non-cooling simulations.  Without
cooling all the relaxed clusters show very similar emission profiles,
with only a small contribution to the total emission coming from the
very centre. These profiles are mostly well resolved, as claimed by Eke
\etal (1998) for simulations of clusters with this particle number.

Once radiative cooling is turned on the radial emission profiles span
a much broader range.  For two of the clusters, a central cooling flow
type emission is clearly visible --- contributing 50 percent and 80
percent of the total X-ray flux.  Although the cooling flow region is
not well-resolved, it cannot be much larger without depleting the
intracluster medium of even more gas.  For each of the other clusters, the
radius enclosing half of the total emission is much larger than that
for the simulation without cooling.  

\subsection{$L_X$--$T_X$ relation}

\begin{figure*}
 \centering
\psfig{file=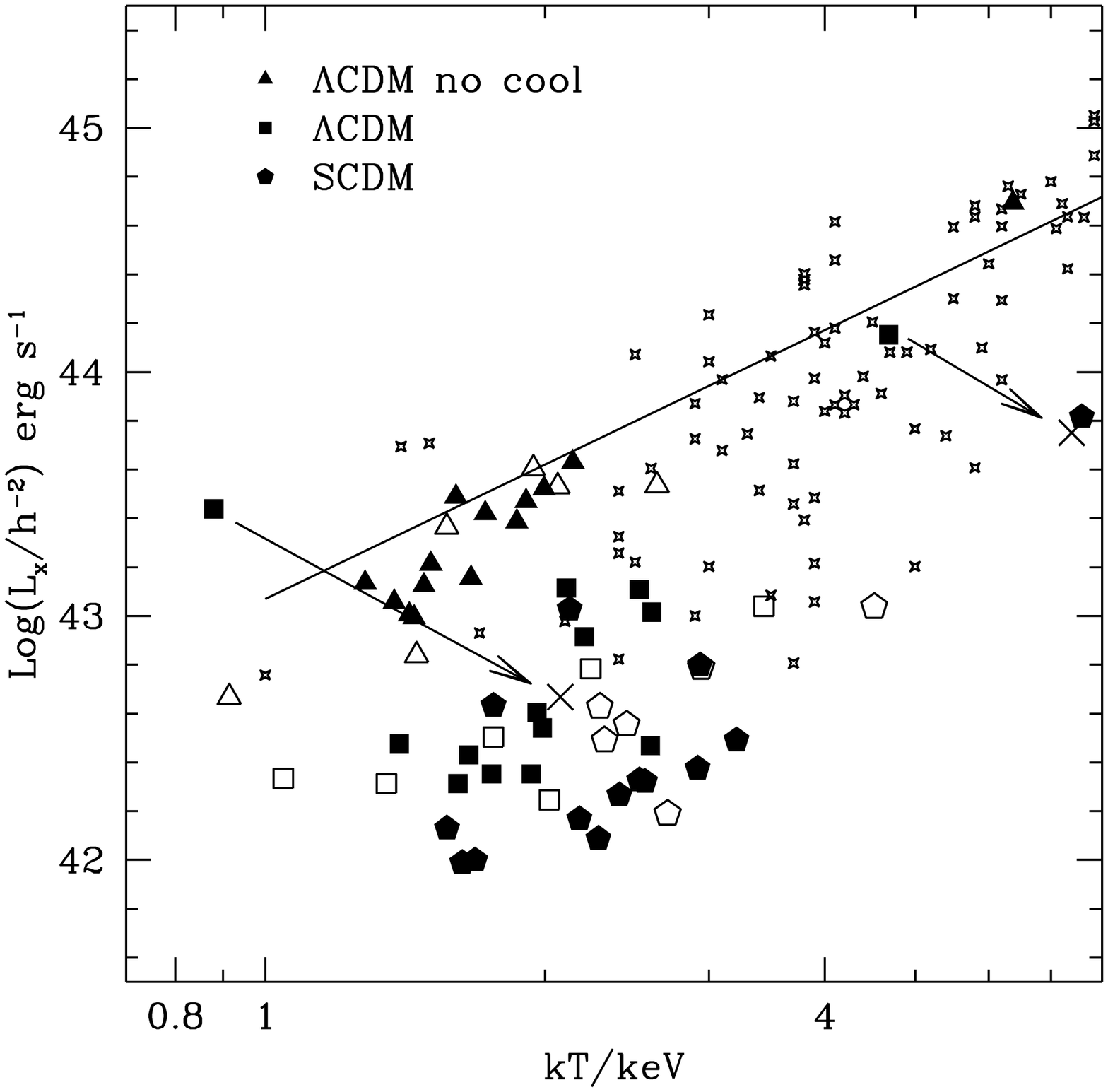,height=15cm}
\caption{The luminosity-temperature relation for all the
clusters. Open symbols refer to those clusters with significant
substructure, filled to those without. The small symbols are the
observation data from David, Jones \& Forman (1995). The majority of 
our clusters are small
because the simulation volume is only $100\Mpc$ on a side
but span a reasonable range. The values plotted are the emission weighted
mean temperature converted to keV and the bolometric luminosity
within the virial radius. The regression line is that of Eke \etal
(1998), Figure 16. and represents a fit to the X-ray luminosity of
their $\Lambda$CDM clusters. The crosses show where the two
$\Lambda$CDM clusters indicated would appear on the diagram if their
central cooling flow emission was removed from both the bolometric
X-ray and emission weighted temperature calculations.}
\label{fig.lxtall}
\end{figure*}

There has been much debate in the literature centering on the X-ray
cluster $L_X$ versus $T_X$ correlation. The emission weighted mean
temperature is plotted against the bolometric luminosity within the
virial radius for all our clusters in Figure~\ref{fig.lxtall}. The
filled symbols represent the relaxed clusters and the open symbols
denote those clusters that show significant substructure.   

The effect of cooling is, in general, to slightly raise the
temperature but to greatly reduce the X-ray luminosity.  Exceptions
are the cooling flow clusters where the large amount of emission from
gas cooling onto the central galaxy gives rise to a lower temperature
than in the non-cooling case.  There are two of these, easily visible
on the plot, in the $\Lambda$CDM run; we have plotted their new locations,
when the central cooling flow is omitted, using crosses linked to
the old location via arrows. In both
cases less than 2 percent of the hot, X-ray emitting particles were 
excised to make this calculation (606 for cluster $\Lambda 1$ and 20
for cluster $\Lambda 5$). In the case of $\Lambda 5$ we have caught a
transient event - a small amount of gas has been reheated by a merger
between a satellite and the brightest cluster galaxy. This gas is in
the process of rapidly cooling back onto the central object, emitting
large amounts of X-rays. 

All 3 sets of clusters display a positive correlation between $L_X$ and
$T_X$, although there are insufficient numbers to tie the trend down
very tightly.  It is clear from the comments in the preceding
paragraph that the nature of the correlation depends critically upon
whether one removes the cooling flow emission or not.  We believe that
a clearer picture arises if this is done.

The regression line in Figure~\ref{fig.lxtall} is from Eke
\etal~(1998) and corresponds to $L_X\propto T_X^2$.  Our non-cooling
clusters fit reasonably well with this relation.  The cooling clusters
lie below this line.  Given that we expect cooling to be less
important in the most massive clusters (the absolute value of the
cooling time and the ratio of the cooling time to the dynamical time
both increase with cluster mass), then we would expect the clusters with
radiative cooling to lie closer to the regression line at higher
$T_X$.  Thus the effect of cooling should be to steepen the
$L_X$--$T_X$ relation.  We hope to test this with more simulations of
higher mass clusters.

Also plotted in Figure~\ref{fig.lxtall} are the observed data from
David \etal (1995). Our clusters are smaller and cooler because they
are not very massive (due to our relatively small computational
volume), thus it is hard to assess whether the cooling or the
non-cooling clusters give a better fit to the data.  SCDM clusters, at
least with the simulation parameters we have chosen in this paper, do
not seem to provide such a good match to the data.

\section{Conclusions and discussion}

We have performed two N-body plus hydrodynamics simulations of
structure formation within a volume of side $100\Mpc$, including the
effects of radiative cooling but neglecting star formation and
feedback. By repeating one of the simulations without radiative
cooling of the gas, we can both compare to previous work and study the
changes caused by the cooling.  A summary of our conclusions follows.

\noindent
(a) Without cooling our clusters closely resemble those found by
previous authors (Eke \etal 1998 and references therein), with dark
matter density profiles that closely follow the universal formula
proposed by Navarro \etal (1995).  

\noindent(b) With cooling, the formation of a central galaxy
within each halo acts to steepen the dark matter profile, supporting
the conclusion of the lensing studies (\eg Kneib \etal 1996) that the
underlying potential that forms the lens has a small core. This galaxy
may not be located exactly at the centre of the X-ray emission,
sometimes being offset by up to $50h^{-1}\kpc$.  The inner slope of
the density profile is then steeper than that suggested by NFW, closer
to that found by Moore \etal (1998) from high resolution N-body
simulations (but note that these did not include cooling gas).

\noindent
(c) We confirm the results of Eke \etal (1998) (and previous studies
quoted therein) that without radiative cooling the gas density profile
turns over at small radii, that the radial temperature profile drops
by a factor of two between its peak value and that obtained at the
virial radius and that the baryon fraction within the virial radius is
lower than the cosmic mean.

\noindent
(d) Cooling acts to remove low-entropy gas from near the cluster
centre, triggering the inflow of higher entropy material.  The entropy
excess compared with the non-cooling run is greatest at about 0.2
times the virial radius, because radiative cooling lowers the entropy
of the gas near the centre of the cluster.

\noindent
(e) We stress the importance of correctly modelling the central
cluster galaxy. The resultant X-ray properties of the cluster are very
dependant upon the centre and if too much material cools into the base
of the potential well large amounts of hot, dense gas can be confined,
producing enormous X-ray fluxes. We have specifically tailored our
models to both globally cool a reasonable fraction of material and to
circumvent problems encountered by SPH when faced with large density
jumps which can lead to high rates of gas cooling onto the central objects.

\noindent
(f) In cooling clusters, the gas density is reduced, by a maximum of
about a factor of three at 0.1--0.2 times the virial radius.  The
density profile more closely resembles a power-law than in the
non-cooling run.  A few clusters show a central density spike,
indicative of a cooling flow onto the central cluster galaxy
(e.g.~Fabian, Nulsen \& Canizares 1991).

\noindent
(g) The temperatures of the cooling clusters show a significant fall
between the point where the peak values are obtained (around 0.1 virial
radii) and the virial radius.  The observational evidence is somewhat
divided here.  Our results are in agreement with Markevitch
\etal (1998) who find that the temperature profile of galaxy clusters are
steeply falling.  However, Irwin \etal (1999) recover isothermal
temperature profiles out to the virial radius from an averaged 
sample of 26 ROSAT clusters. 
It is important to try to clear up this observational controversy 
as isothermal temperature profiles are not seen in our models but
are often used for X-ray mass estimates,
measurements of $\Omega_b$ (and hence $\Omega$) and theoretical
arguments for the $L_X$--$T_X$ relation.

\noindent
(h) Cooling acts so as to \emph{increase} both the mass-weighted and
observed X-ray temperatures of clusters.  The peak temperatures are
raised by a factor of about 1.5 and the temperature gradient between
the peak and the virial radius is correspondingly increased.

\noindent
(i) The bolometric luminosity for the clusters with radiative cooling
is around 3--5 times \emph{lower} than for matching clusters without it.
Except for the cooling flow clusters, the X-ray luminosity profile is
less centrally concentrated than in the non-cooling case with a
greater contribution coming from larger radii.

\noindent
(j) Cooling flow clusters are easily distinguished from non-cooling
flow clusters in the $L_X$--$T_X$ plane.  The former are  more
luminous and cooler than the latter (Fabian et al 1994,
Allen \& Fabian 1998).  Some of this difference results from
X-ray analysis methods (Allen \& Fabian 1998) but may also be caused by
actual physical differences in the mass distribution
of clusters. We suggest that, while interpretation of this 
relation would be made simpler if the
cooling flow were excluded before determining the X-ray properties of
clusters, this should be done with caution.

\noindent
(k) The clusters from the non-cooling simulation lie on the
$L_X\propto T_X^2$ regression line of Eke \etal~(1988), whereas those
from the cooling run lie below it.  We suggest that, as cooling is
likely to be less important in more massive clusters, the effect of
cooling will be to steepen the relation.  This remains to be tested
with simulations of more massive clusters.

\noindent
(l) The very large core radii ($\sim 250h^{-1}\kpc$) 
observed in some clusters are not seen in our simulations.
It is possible that such events are rare, or occur preferentially in
massive clusters, in which case our cluster sample may simply be too
small to contain such objects.  Smaller core radii of around $\sim
50\,h^{-1}\kpc$ are close to our resolution limit and so are not ruled
out by our results, although we find no evidence for a
constant-density region in the centre of any of our clusters.

At this point, we should remind the reader that our simulations are
designed merely to investigate the effect of radiative cooling on the
X-ray properties of the intracluster medium and do not set out to
explore the complete range of physical processes going on in clusters.
In particular, we ignore the possibility of energy injection into the
intracluster medium.  
Early star formation at high redshift and the subsequent
energy feedback from supernovae could act to preheat the gas which
falls into the potential wells of galaxy clusters, effectively raising
its entropy to such a level that it cannot reside at the high
densities required to trace the dark matter into the base of the
potential well, and giving rise to the constant-density cores that our
simulated clusters lack.  Ponman, Cannon
\& Navarro (1999) argue for such preheating by examining ROSAT
observations of 25 clusters.

The inclusion of cooling into a cosmological hydrodynamics simulation
has proved highly successful.  We have highlighted the necessity of
cooling reasonable amounts of gas onto the centre of each galaxy
cluster if sensible X-ray luminosity estimates are to be obtained.
We have achieved this by a judicious choice of mass resolution and
decoupling the hot and cold gas phases.  
A rigorous method of doing
this in SPH is under development.  We next intend to simulate more
massive clusters in the same way, in order to extend our predictions
into regions more accessible to observation.

\section*{Acknowledgments}

The work presented in this paper was carried out as part of the
programme of the Virgo Supercomputing Consortium using computers based
at the Computing Centre of the Max-Planck Society in Garching and at
the Edinburgh Parallel Computing Centre.  The authors thank NATO for
providing NATO Collaborative Research Grant CRG 970081 which
facilitated their interaction and the anonymous referee for suggesting
significant improvements.  PAT is a PPARC Lecturer Fellow.

\end{document}